\documentclass[a4paper,11pt]{article}
\pdfoutput=1 

\usepackage{jinstpub} 

\title{Time imaging reconstruction for the PANDA Barrel DIRC}

\author[a,1]{R.~Dzhygadlo,\note{Corresponding author.}}
\author[a,b]{A.~Ali,}
\author[a]{A.~Belias,}
\author[a]{A.~Gerhardt,}
\author[a]{M.~Krebs,}
\author[a]{D.~Lehmann,}
\author[a,b]{K.~Peters,}
\author[a]{G.~Schepers,}
\author[a]{C.~Schwarz,}
\author[a]{J.~Schwiening,}
\author[a]{M.~Traxler,}
\author[c]{L.~Schmitt,}
\author[d]{M.~B\"{o}hm,}
\author[d]{A.~Lehmann,}
\author[d]{M.~Pfaffinger,}
\author[d]{S.~Stelter,}
\author[d]{F.~Uhlig,}
\author[e]{M.~D\"{u}ren,}
\author[e]{E.~Etzelm\"{u}ller,}
\author[e]{K.~F\"{o}hl,}
\author[e]{A.~Hayrapetyan,}
\author[a,e]{I.~K\"{o}seoglu,}
\author[e]{K.~Kreutzfeld,}
\author[e]{J.~Rieke,}
\author[e]{M.~Schmidt,}
\author[e]{T.~Wasem,}
\author[f]{C.~Sfienti}

\affiliation[a]{GSI Helmholtzzentrum f\"ur Schwerionenforschung GmbH, Darmstadt, Germany}
\affiliation[b]{Goethe University, Frankfurt a.M., Germany}
\affiliation[c]{FAIR, Facility for Antiproton and Ion Research in Europe, Darmstadt, Germany}
\affiliation[d]{Friedrich Alexander-University of Erlangen-Nuremberg, Erlangen, Germany}
\affiliation[e]{II. Physikalisches Institut, Justus Liebig-University of Giessen, Giessen, Germany}
\affiliation[f]{Institut f\"{u}r Kernphysik, Johannes Gutenberg-University of Mainz, Mainz, Germany}

\emailAdd{r.dzhygadlo@gsi.de}

\abstract{The innovative Barrel DIRC (Detection of Internally Reflected Cherenkov light)
  counter will provide hadronic particle identification (PID)
  in the central region of the PANDA experiment at the new Facility
  for Antiproton and Ion Research (FAIR), Darmstadt, Germany.
  This detector is designed to separate charged pions and kaons
  with at least 3 standard deviations for momenta up to 3.5 GeV/c,
  covering the polar angle range of $22^{\circ}-140^{\circ}$.
  An array of microchannel plate photomultiplier tubes is used to
  detect the location and arrival time of the Cherenkov photons with a
  position resolution of 2 mm and time precision of about 100 ps.
  The time imaging reconstruction has been developed to make
  optimum use of the observables and to determine
  the performance of the detector. This reconstruction algorithm performs
  particle identification by directly calculating the maximum likelihoods
  using probability density functions based on detected photon propagation time
  in each pixel, determined directly from the data, or analytically, or
  from detailed simulations.
}

\keywords{Cherenkov detectors; Particle identification methods.}

\begin{document}
\maketitle
\flushbottom

\section{Introduction}
\label{sec:intro}

The PANDA Barrel DIRC \cite{barrel_dirc,barrel} is a key component of the particle
identification (PID) system for the PANDA detector \cite{panda_tech},
which will be installed at the Facility for Antiproton and Ion Research (FAIR) in Germany. 
The PID goal for the Barrel DIRC is to reach 3 standard deviations (s.d.) $\pi/K$ separation
for momenta up to 3.5 GeV/c, covering the polar angle range of $22^{\circ}-140^{\circ}$.

\begin{figure}[htbp]
  \centering
  \includegraphics[width=.98\textwidth]{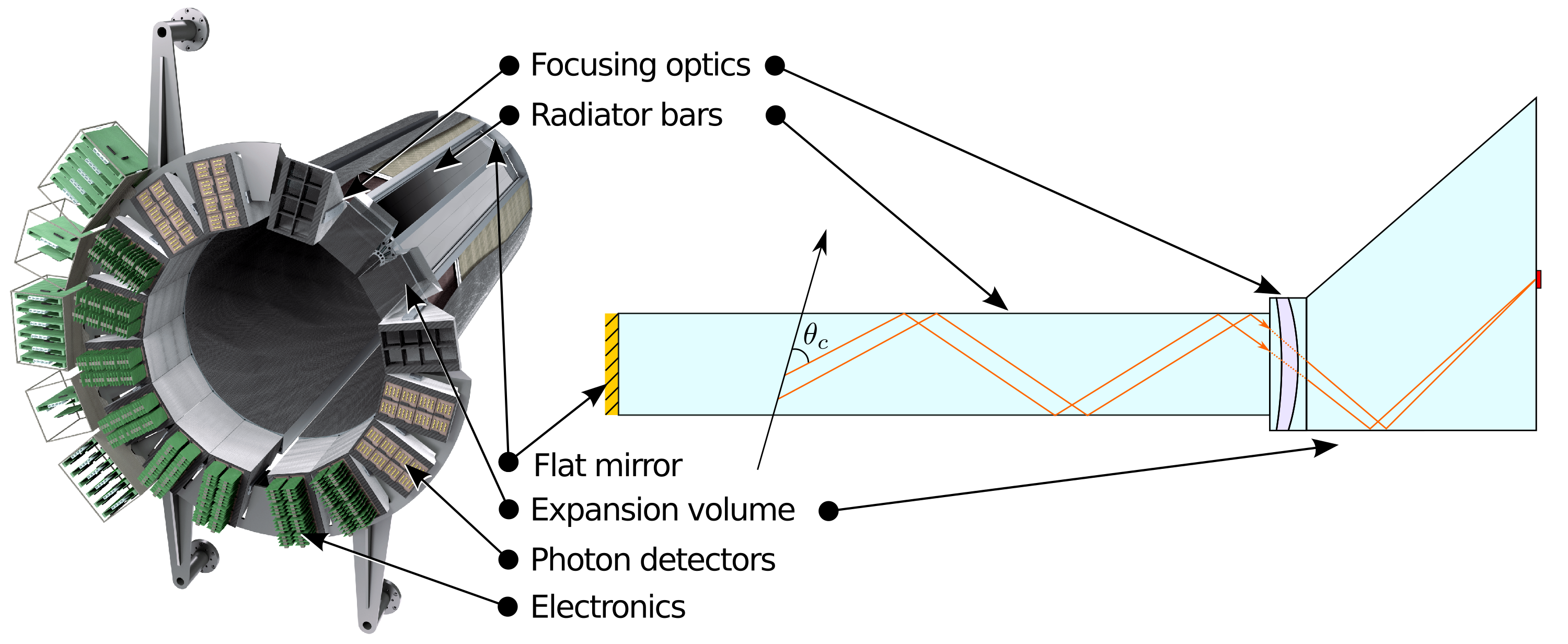}
  \caption{\label{fig:dirc} Rendered CAD drawing of the PANDA Barrel DIRC (left)
    and the simplified cross section of one Barrel DIRC sector (right, not to scale).}
\end{figure}

The Barrel DIRC is constructed in the shape of a barrel using 16
optically isolated sectors, each comprising
a radiator box and a compact, prism-shaped expansion volume (EV)
(see figure~\ref{fig:dirc}). 
The radiator box contains three synthetic fused silica bars of
17 $\times$ 53 $\times$ 2400~mm$^3$ size,
positioned side-by-side with a small air gap between them.
A flat mirror at the forward end of each bar is used to reflect
Cherenkov photons to the read-out end, where a 3-layer
spherical lens images them on an array of 8 Microchannel Plate
Photomultiplier Tubes (MCP-PMTs).
The MCP-PMT has 64 pixels of 6.5~$\times$~6.5~mm$^2$ size
and, in combination with the FPGA-based readout electronics,
will be able to detect single photons with a precision of about $100$~ps.

\begin{figure}[htbp]
  \hspace*{0.5cm}
  \includegraphics[width=.47\textwidth]{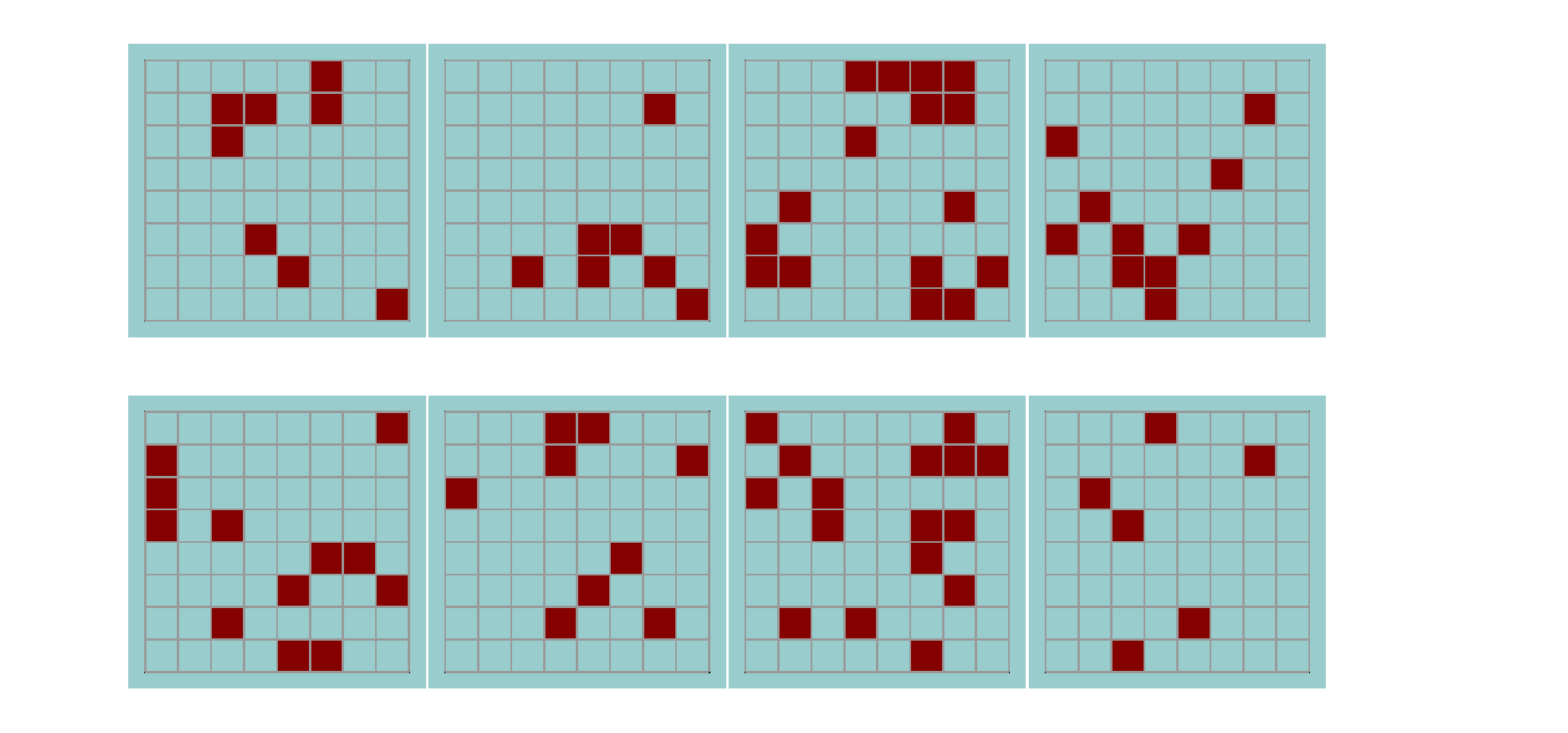}
  \includegraphics[width=.47\textwidth]{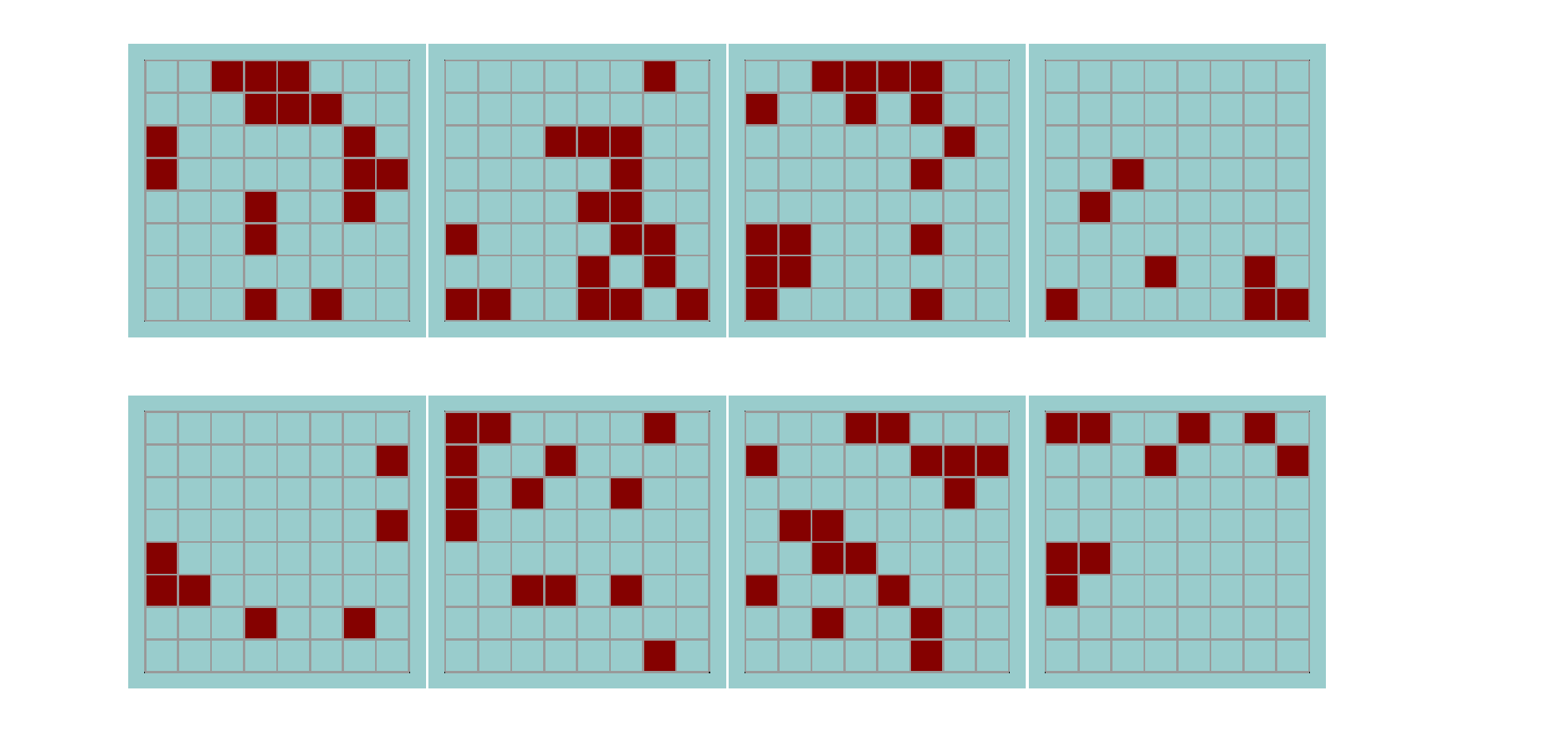}
  \qquad
  \hspace*{0.5cm}
  \includegraphics[width=.45\textwidth]{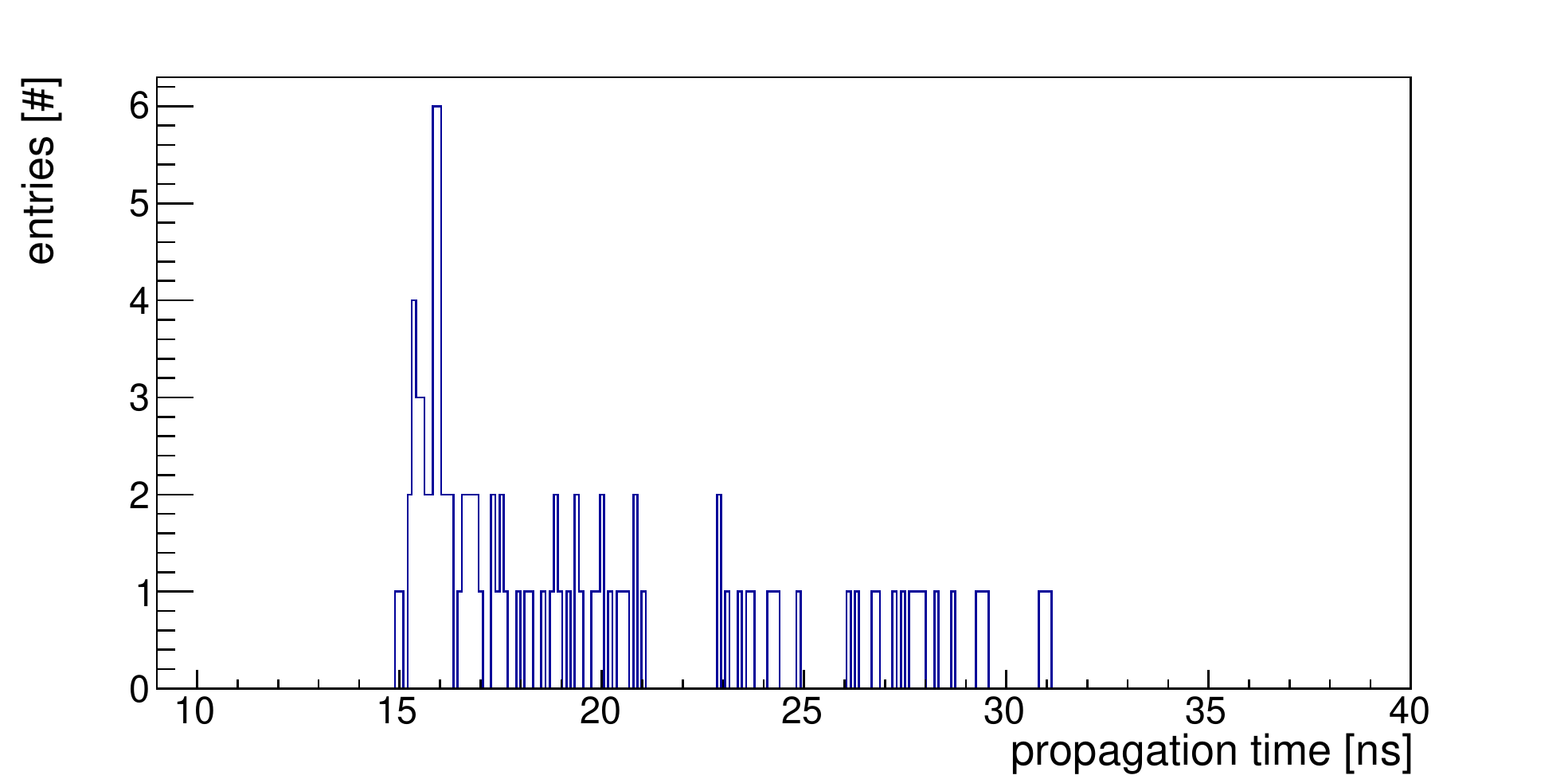}
  \hspace*{0.3cm}
  \includegraphics[width=.45\textwidth]{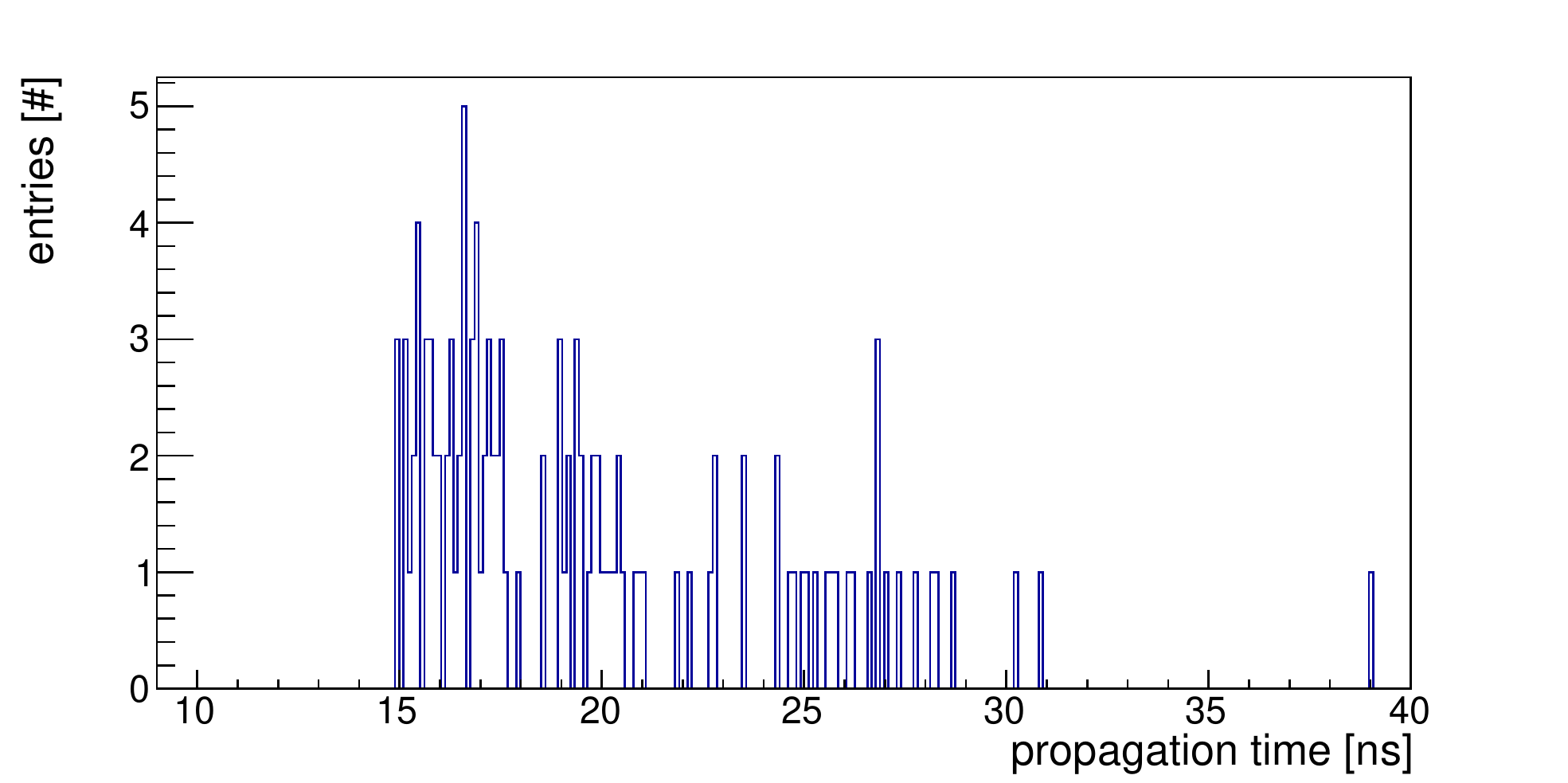}
  \caption{\label{fig:observables} Hit patterns (top) and time spectra (bottom)
    for a single pion (left) and kaon (right) at $22^{\circ}$ polar angle
    and 3.5 GeV/c momentum.}
\end{figure}

Depending on the polar angle and momentum of the charged particle,
the system detects 20-100 photons.
Figure~\ref{fig:observables} shows a typical hit pattern and
time spectra for a single pion (left) and kaon (right) at $22^{\circ}$
polar angle and 3.5 GeV/c momentum. Using this information in
combination with knowledge of the charged particle momentum and direction,
the reconstruction algorithms perform particle identification (PID).
Two algorithms have been developed to make optimum use
of the observables and to determine the performance of the detector.
The "geometrical reconstruction" \cite{barrel_reco}, initially developed
for the BaBar DIRC \cite{babar}, performs PID by reconstructing
the value of the Cherenkov angle and using it in a track-by-track
maximum likelihood fit, relying mostly on the
position of the detected photons in the reconstruction, using the time
information primarily to suppress backgrounds.
The "time imaging" utilizes both, position and time information,
and directly performs the maximum likelihood fit.

\section{Time Imaging Reconstruction}

\begin{figure}[htbp]
  \includegraphics[width=.498\textwidth]{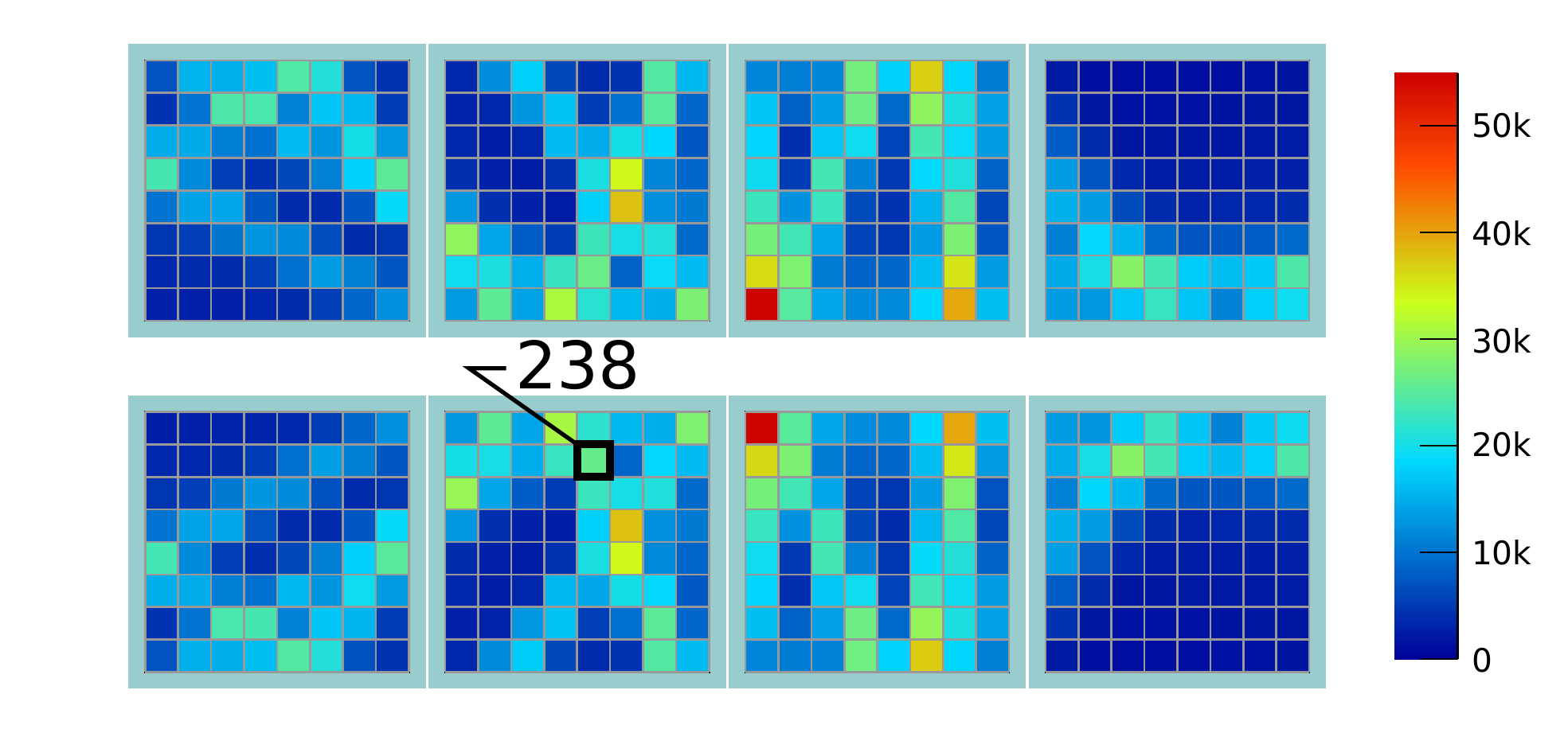}
  \includegraphics[width=.498\textwidth]{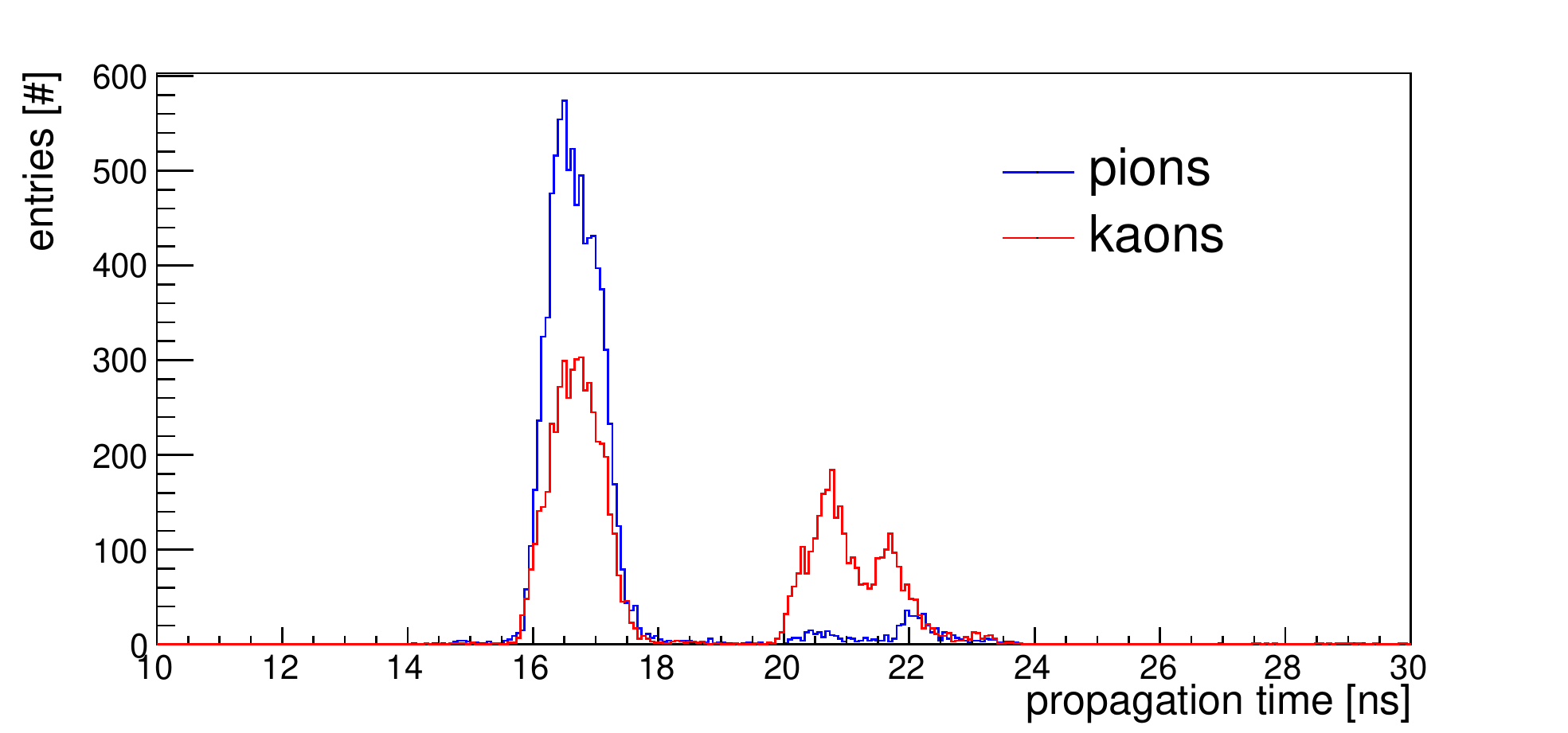}
  \caption{\label{fig:accumulated} Accumulated hit pattern (left) and
    the propagation time spectra for one example pixel, number 238 (right), for 30k pions and
    kaons simulated at $22^{\circ}$ polar angle and 3.5 GeV/c momentum.}
\end{figure}

The time imaging method is based on the approach used by the
Belle II time-of-propagation (TOP) counter \cite{staric_2}. The basic concept
is that the measured arrival time of Cherenkov photons in
each single event is compared to the expected photon
arrival time for every pixel and for every particle
hypothesis, yielding the PID likelihoods. Figure~\ref{fig:accumulated}
shows an example of the accumulated hit pattern and the propagation time spectra
for 30k simulated pions and kaons for one specific pixel. The arrival time
of the Cherenkov photons produced by  $e$, $\mu$, $\pi$, K, and p
is normalized for every pixel to produce probability
density functions (PDFs). The total PID likelihood is then calculated as:
\begin{equation} \label{eq:lh}
  \log\mathcal{L}_h = \displaystyle\sum^{N}_{i=1}\log\big(S_h(p_i,t_i)+B(p_i)\big)+\log P_h(N),
\end{equation}
where $N$ is the number of detected photons in a given event,
$S_h(p_i,t_i)$ is the PDF for a pixel $p_i$
and particle type $h$, and $B(p_i)$ is the expected background contribution,
which includes MCP-PMT dark noise and accelerator background. 

The second term in Eq.~\ref{eq:lh} is the Poisson distribution, which
accounts for a difference in photon yields of different particle types.
This contribution can be quite significant at low momenta but is negligible at
higher momenta, where the photon yield is almost independent of the particle type.

\section{Probability Density Functions}

The PDFs are created from the photon arrival time, which can
be obtained in several ways. The best PID performance is expected
from the PDFs created using propagation times from the experimental data. In this case,
the propagation time is a direct measurement which already includes all
detector imperfections and, therefore, provides the most realistic PDFs. In this method
a large amount of data for the whole angular and momentum acceptance is required.
If the amount of experimental data is not sufficient, a full detector simulation can
be used to pre-generate a large number of tracks. Both methods require
a large amount of memory to store all possible PDFs and, therefore, are not
practical for application in PANDA. The full simulation can also be performed
during reconstruction for each event with a given track
direction but excessive simulation time makes it, again, impractical to use.
\begin{figure}[htbp]
  \centering  
  \includegraphics[width=.54\textwidth]{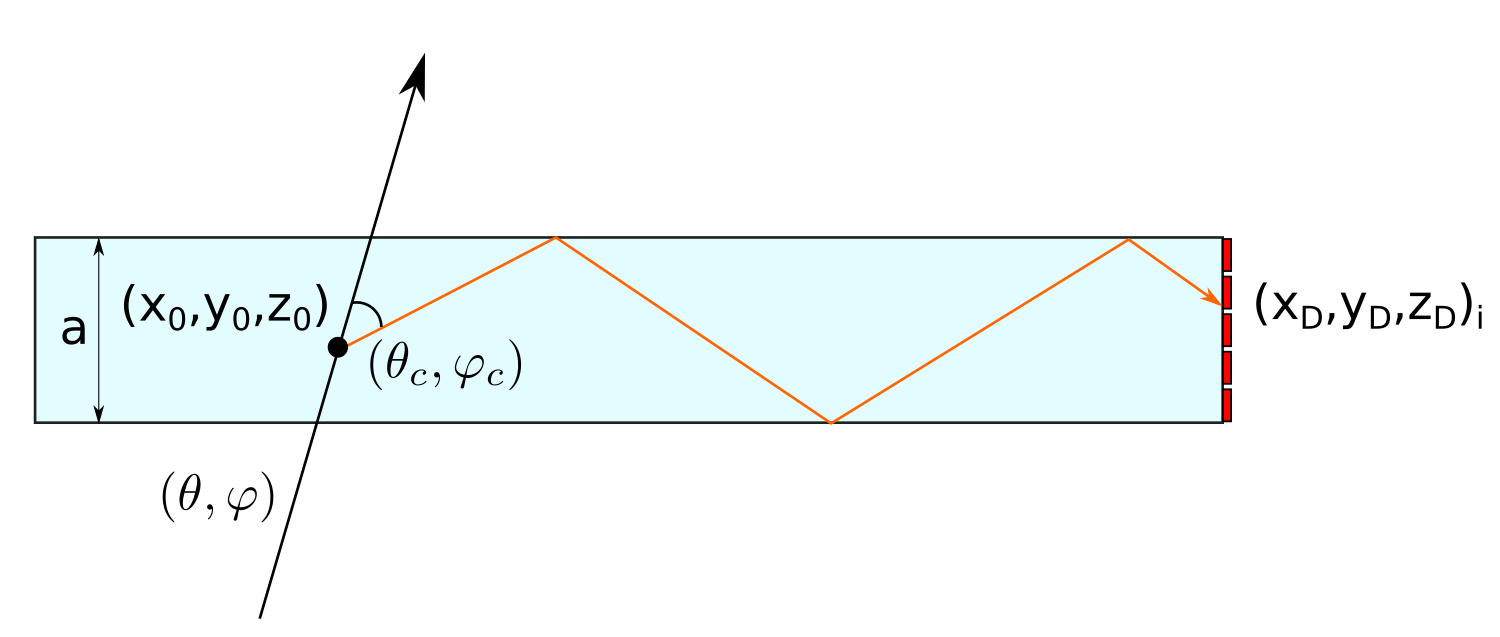}
  \includegraphics[width=.45\textwidth]{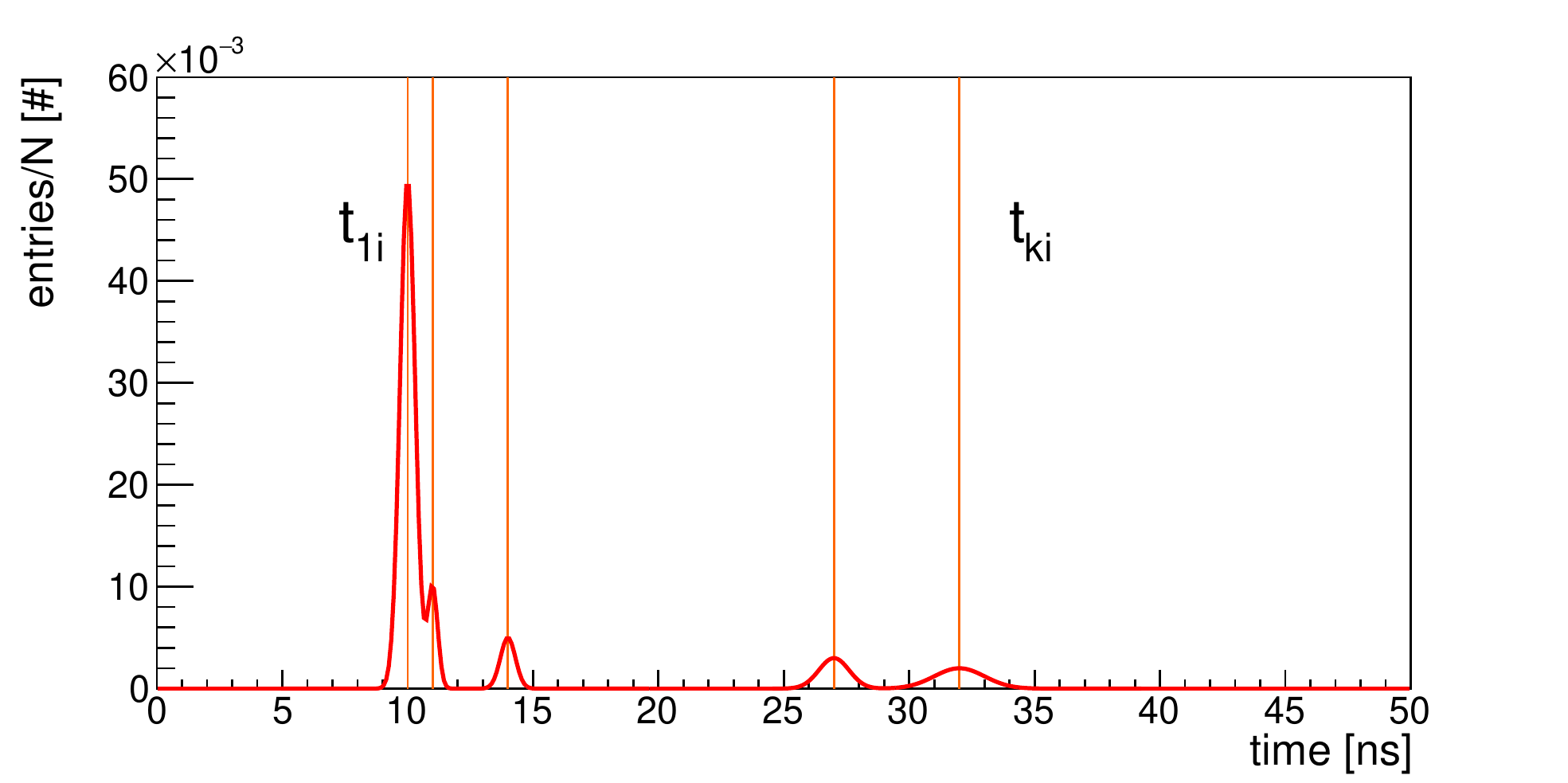}
  \caption{\label{fig:pdf_simple} Simplified detector configuration without expansion
    volume and focusing system (left). Example of a PDF as a superposition of Gaussians
    with mean values $t_{ki}$ (right).}
\end{figure}
Finally, the PDFs can be calculated analytically, as shown
by the Belle~II~TOP group \cite{staric_2}. 
In this case, the PDF $S_h$ is represented as a sum of $m_i$ weighted Gaussians:
\begin{equation} \label{eq:gauss}
  S_h(p_i,t_i) = \sum^{m_i}_{k=0}n_{ki}g(t_{ki},\sigma_{ki}),
\end{equation}
where $n_{ki}$ is the number of photons in the $k$-th peak of the pixel $i$, $t_{ki}$
and $\sigma_{ki}$ are position and width of the peak, respectively.

Considering a simplified configuration without expansion volume and without focusing system
(see figure~\ref{fig:pdf_simple}, left), the positions of the Gaussian peaks
can be expressed through the direction of the charged track ($\theta,\varphi$), the
Cherenkov angle $\theta_c$ of the assumed particle hypothesis, and the
positions of the emission ($x_0,y_0,z_0$) and detection ($x_d,y_d,z_d$)
of the Cherenkov photons:
\begin{equation} \label{eq:sol_1}
  \begin{split}
    &t_{ki} = \frac{z_d-z_0}{\big(\cos\theta\cos\theta_c - \sin\theta\sin\theta_c\cos\phi^{ki}_c \big)} \frac{n_g}{c_0},
  \end{split}
\end{equation}
where $n_g$ is the group refractive index of the radiator,
$c_0$ is the speed of light in vacuum,
and $\phi^{ki}_c$ is the azimuthal angle of the Cherenkov photon
in the charged particle's coordinate system, which is defined as:
\begin{equation} \label{eq:sol_2}
  \begin{split}
    &\cos\phi^{ki}_c = \frac{a_{ki}b_{ki} \pm d\sqrt{d^2+b^2_{ki}-a^2_{ki}}}{b^2_{ki}+d^2},
  \end{split}
\end{equation}
where
\begin{equation} \label{eq:sol_3}
  \begin{split}
    &a_{ki} = \frac{x^{ki}_d-x_0}{z_d-z_0}\cos\theta\cos\theta_c - \cos\varphi\sin\theta\cos\theta_c,\\
    &b_{ki} = \frac{x^{ki}_d-x_0}{z_d-z_0}\sin\theta\sin\theta_c + \cos\varphi\cos\theta\sin\theta_c,\\
    &d = \sin\varphi\sin\theta_c.
  \end{split}
\end{equation}
Here the value of $x^{ki}_d$ represents the exit position of the Cherenkov
photon in the unfolded radiator plane at $z_d$:
\begin{equation} \label{eq:k}
  x^{ki}_d = \begin{cases}
    ka + x^{i}_d, & k = 0,\pm2,\pm4, ... \\
    ka - x^{i}_d, & k = \pm1,\pm3, ...,
  \end{cases}      
\end{equation}
where $k$ is the number of reflections inside the radiator and is a running parameter.

The width $\sigma_{ki}$ includes
contributions from the photon emission spread $\Delta\lambda$,
multiple scattering $\Delta\theta$, chromatic error $\Delta\sigma_e$,
pixel size $\Delta x^{i}_d$ and the propagation time measurement error $\sigma_{\mathrm{m}}$: 
\begin{equation} \label{eq:sol_4}
  \sigma_{ki} = \sqrt{\left(\frac{\partial t_{ki}}{\partial \lambda} \right)^2 \Delta\lambda^2+\left(\frac{\partial t_{ki}}{\partial \theta} \right)^2 \Delta\theta^2+\left(\frac{\partial t_{ki}}{\partial \sigma_e} \right)^2 \Delta\sigma^2_e+\left(\frac{\partial t_{ki}}{\partial x^{i}_d} \right)^2 \Delta x^{i2}_d + \sigma^{2}_{\mathrm{m}}}.
\end{equation}
Finally, the number of photons in each peak is defined as:
\begin{equation} \label{eq:sol_5}
  n_{ki} = N_0l\sin^2\theta_c\frac{\Delta\phi^{ki}_c}{2\pi},
\end{equation}
where $N_0$ is the Cherenkov photon production constant,
$l$ is the length of the charged particle
trajectory in the radiator, and $\Delta\phi^{ki}_c$ is the range of the
Cherenkov azhimuthal angle coverage of the $i$-th pixel.

\begin{figure}[htbp]
  \includegraphics[width=.498\textwidth]{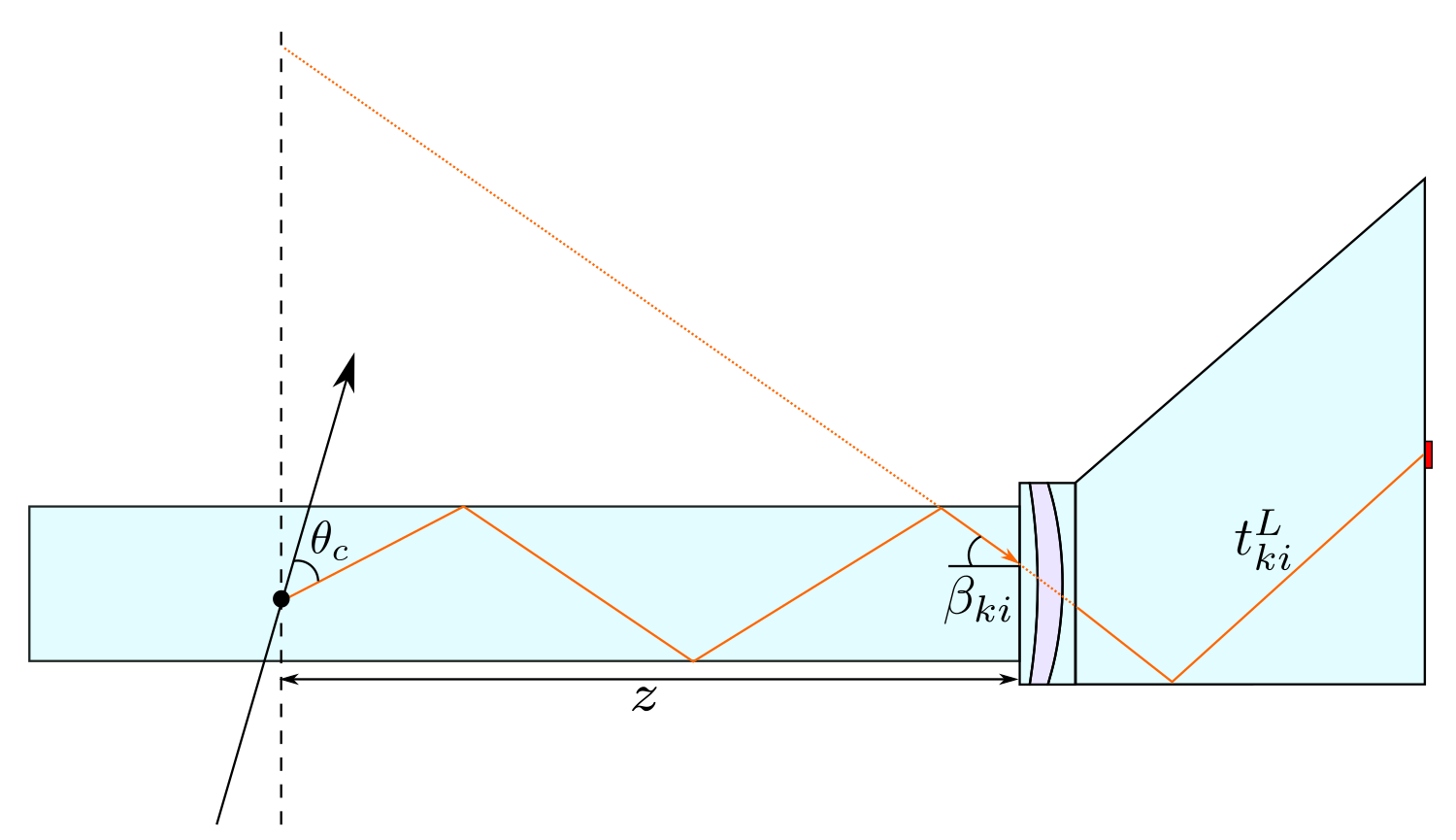}
  \includegraphics[width=.498\textwidth]{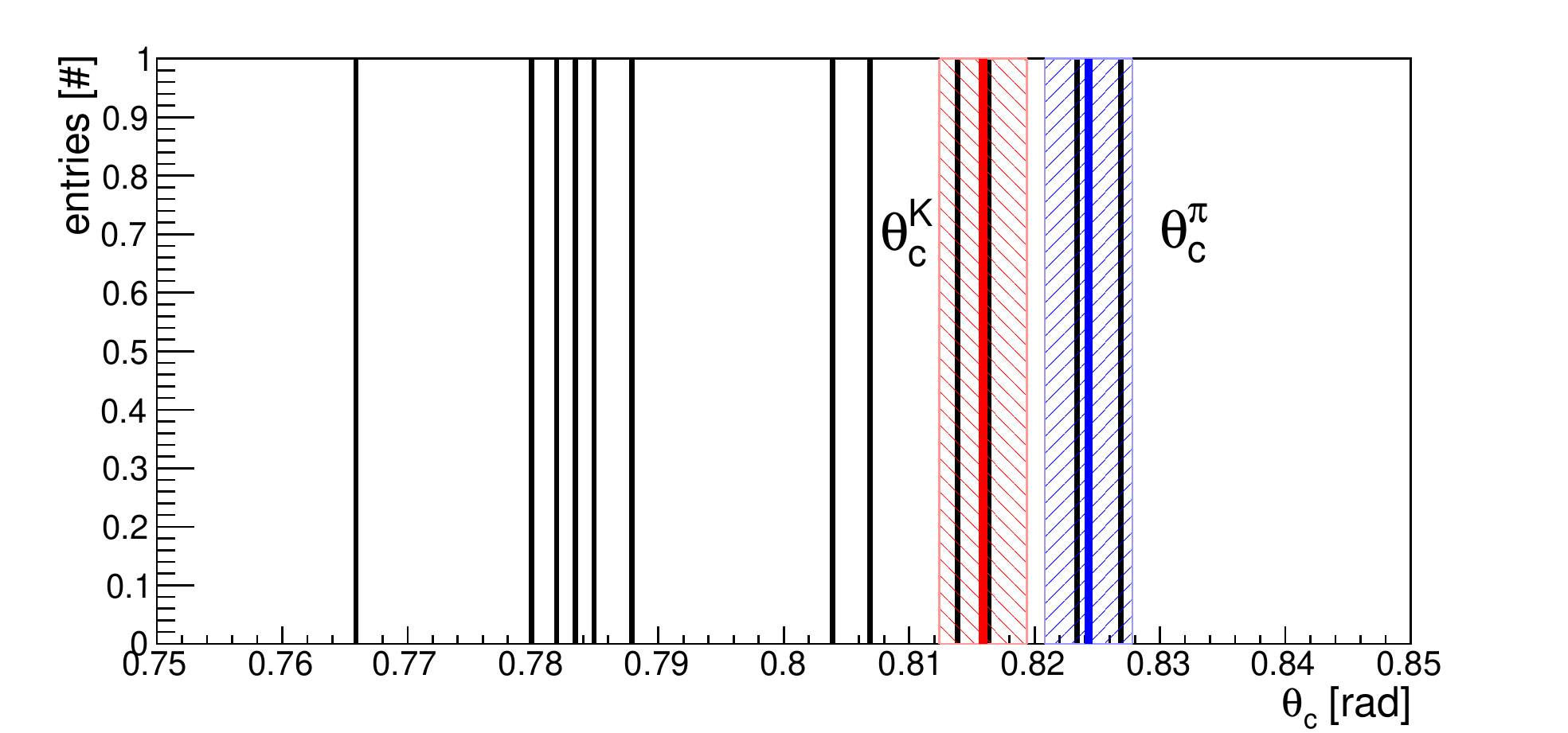}
  \caption{\label{fig:lut} Detector configuration with an expansion volume
    and focusing (left). Example of the LUT solutions with a selection based on the
    reconstructed Cherenkov angle (right). Vertical black lines show the values
    of the reconstructed Cherenkov angle for one detected photon. 
    Shaded red and blue areas show the selection around
    the expected Cherenkov angle for kaons and pions, respectively.}
\end{figure}

By adding the expansion volume and focusing system, the positions of the
photons exiting the radiator $(x_d,y_d,z_d)_i$ become ambiguous.
An additional running parameter can be used to mitigate this but
it will significantly slow down the reconstruction speed.
Instead, a look up table (LUT) is used to determine the exit direction
of the Cherenkov photon from the radiator. The LUT is constructed using
Geant4 \cite{geant4} simulations and comprises all possible directions
from the end of radiator which can lead to a hit in a given pixel.
The Gaussian mean $t_{ki}$ then can be determined as
(see also figure~\ref{fig:lut}, left):
\begin{equation}  \label{eq:lut}
  t_{ki} = \frac{z}{\cos\beta_{ki}}\frac{n_g}{c_0}+t^{L}_{ki},
\end{equation}
where $z$ is the distance from the photon emission point to the readout end
of the radiator and  $t^{L}_{ki}$ is the propagation time of the photon inside the
expansion volume, which is also stored in the LUT. 

The tagging of the determined Gaussian peaks $g(t_{ki},\sigma_{ki})$
with a particle hypothesis is done by reconstructing the Cherenkov angle $\theta^{\mathrm{LUT}}_c$
using the geometrical method \cite{barrel_reco} and comparing it to the expected
value of the given particle hypothesis $\theta^h_c$ (see also figure~\ref{fig:lut}, right):
\begin{equation}  \label{eq:sel}
  \left|\theta^{h}_{c}-\theta^{\mathrm{LUT}}_c \right|<w\sigma_{\mathrm{SPR}},
\end{equation}
where $\sigma_{\mathrm{SPR}}$ is the single photon resolution of the DIRC counter,
and $w$ is the selection constant, which varies in a range of 0.3-1 depending on
the polar angle of the charged particle.

The construction of the PDF $S_h$ for a given particle hypothesis
is done using Eq.~\ref{eq:gauss} with $t_{ki}$ from Eq.~\ref{eq:lut} which
survives the Cherenkov angle selection Eq.~\ref{eq:sel}.

\section{Performance}

\begin{figure}[htbp]
  \centering  
  \includegraphics[width=.75\textwidth]{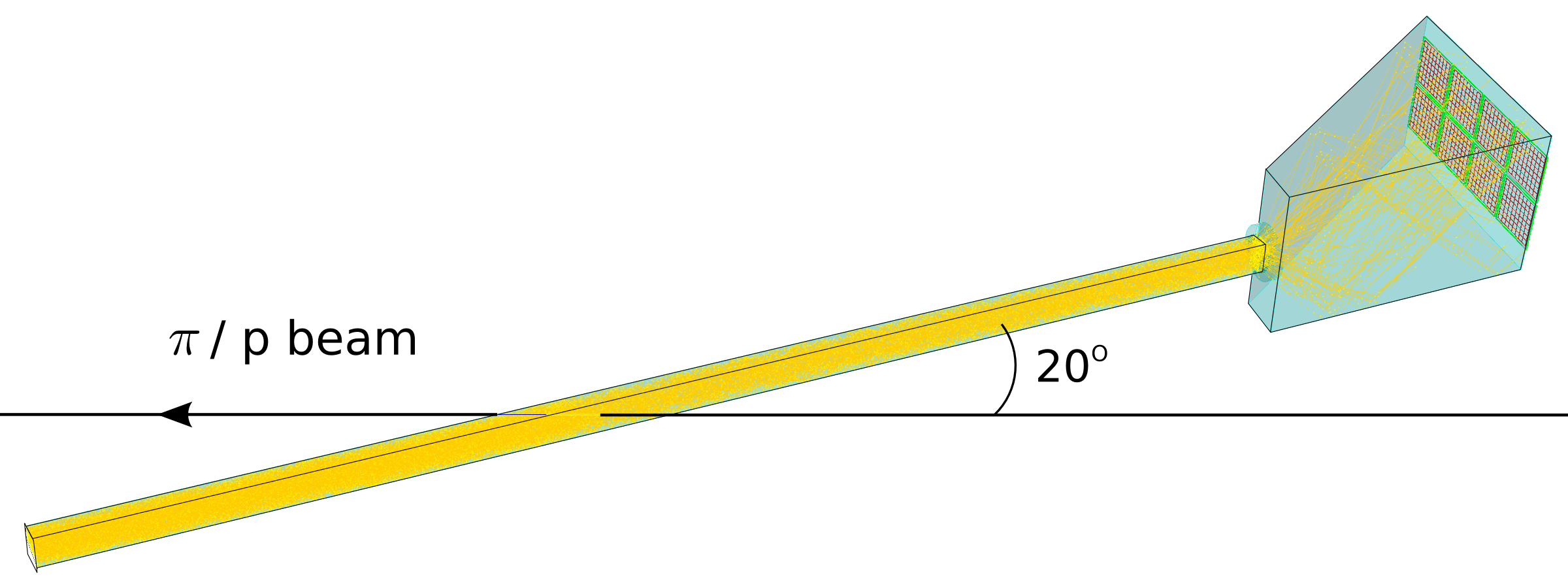}
  \caption{\label{fig:prt} Geant4 simulation of the DIRC prototype
    configuration at $20^{\circ}$ polar angle.
    Yellow lines show the path of Cherenkov photons inside the bar and the prism.}
\end{figure}

The performance of the algorithm was evaluated with Geant4 simulation of the
prototype configuration which was tested with a $\pi/p$ beam at CERN PS
in 2018 \cite{cern18} (see figure \ref{fig:prt}).
The Barrel DIRC prototype contained all relevant parts of one
PANDA Barrel DIRC sector. A narrow fused silica bar
(17.1 $\times$ 35.9 $\times$ 1200.0~mm$^3$) was used as radiator.
It coupled on one end to a flat mirror, on the other end to
a 3-layer spherical focusing lens with a fused silica prism as EV.
An array of 2$\times$4 MCP-PMTs attached to the back side of
the EV was used to detect Cherenkov photons.
The momentum of the mixed hadron beam was set to 7 GeV/c since
$\pi/p$ PID challenge at this momenta is equivalent to $\pi/K$'s at 3.5~GeV/c,
due to similar Cherenkov angle difference.
A time-of-flight system was used to cleanly tag pions and protons.

\begin{figure}[htbp]
  \includegraphics[width=.498\textwidth]{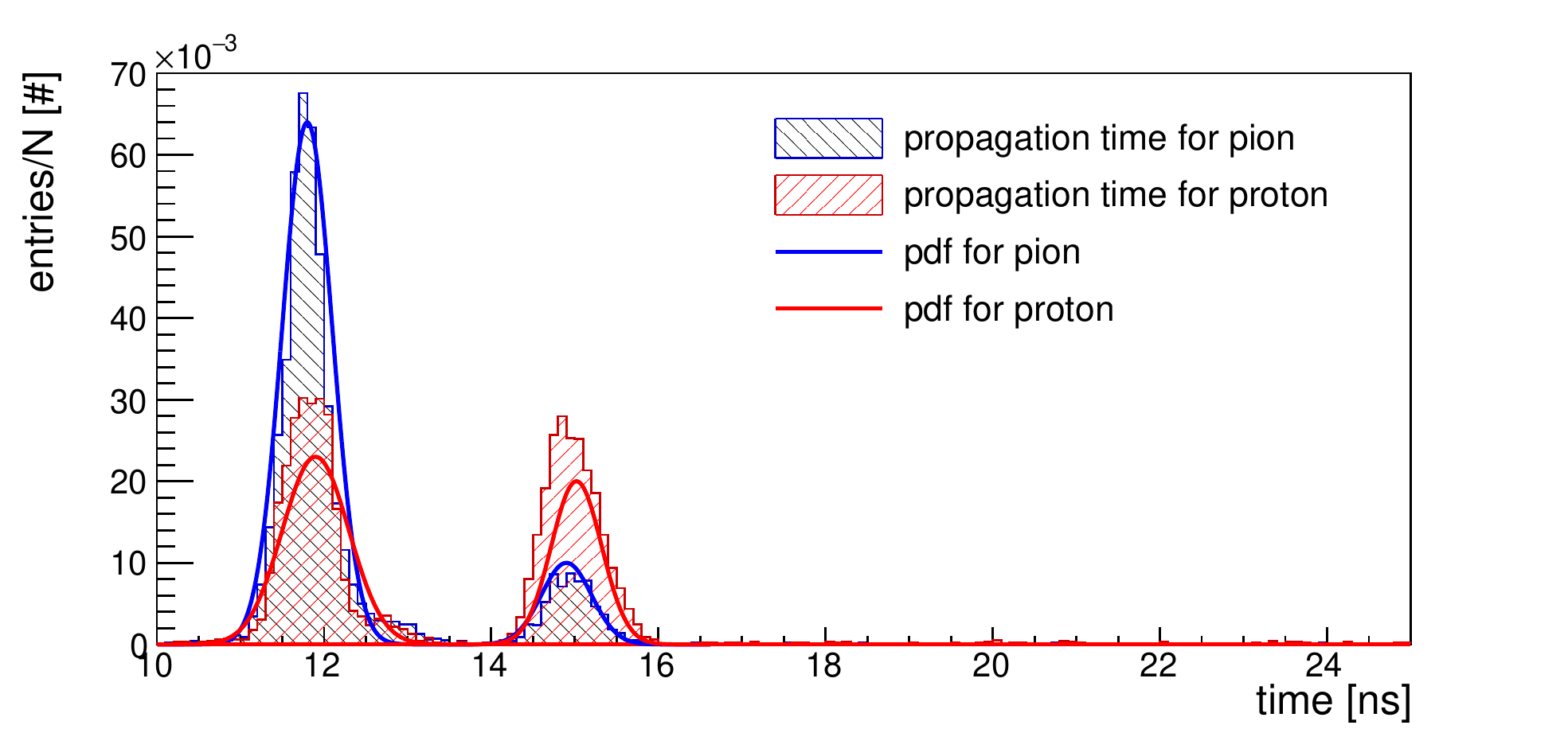}
  \includegraphics[width=.498\textwidth]{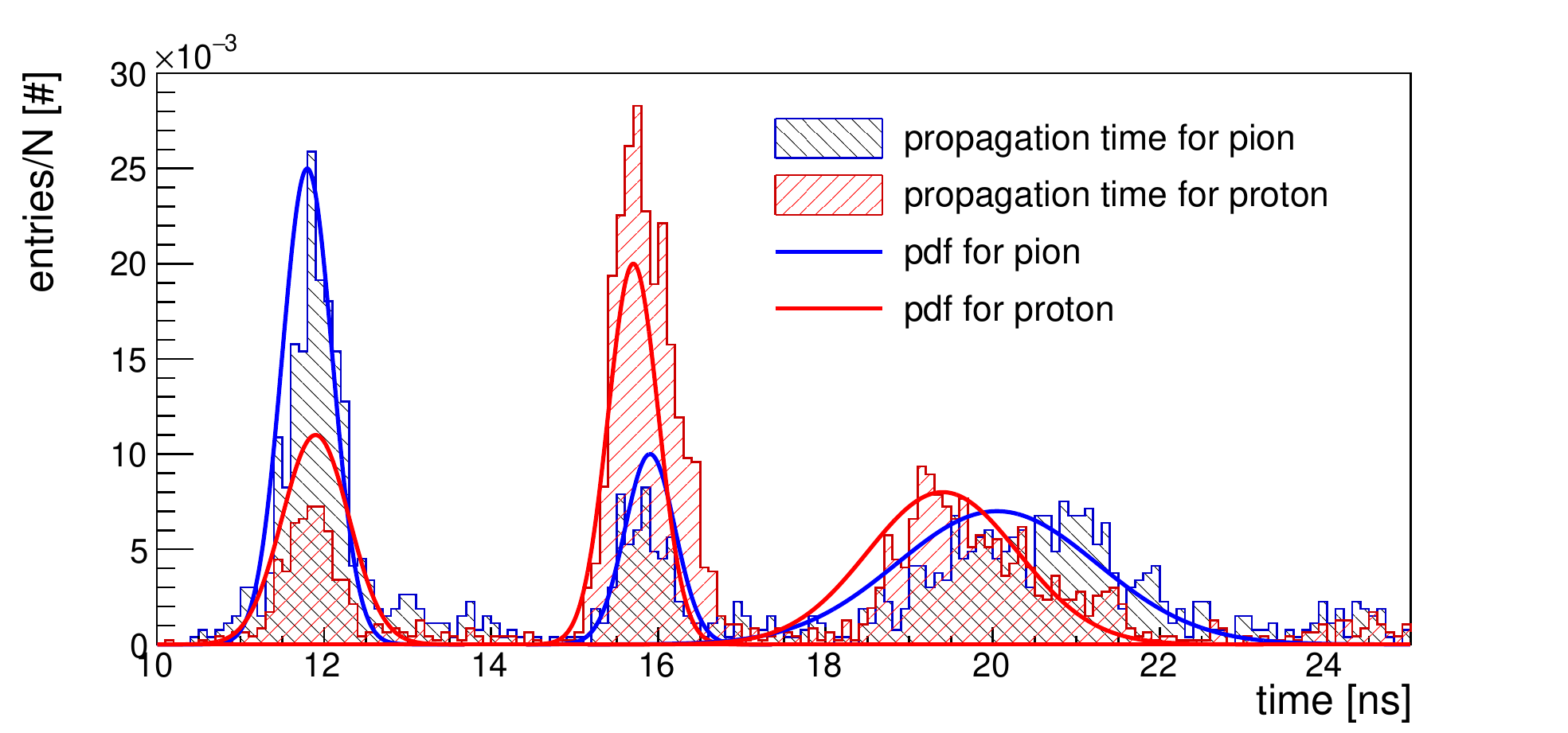}
  \caption{\label{fig:pdf} Examples of PDFs for two pixels, number 245 (left)
    and 285 (right) for pions (blue) and protons (red) at $20^{\circ}$ polar
    angle and 7 GeV/c momentum.
    Shaded histograms correspond to the Geant4 simulations for 30k pions and protons while
    the solid colored lines show analytically determined PDFs.}
\end{figure}

Figure~\ref{fig:pdf} shows an example of analytical PDFs (solid lines) compared
to simulated distributions (shaded histograms) for 30k pions (blue) and protons (red)
at $20^{\circ}$ polar angle.
The analytical PDFs were obtained with selection constant $w$=0.5 and are
in a reasonable agreement with simulation.
A slight disagreement in the heights and the positions of the peaks is the result
of using idealized geometry for creation of the analytical PDF.

\begin{figure}[htbp]
  \includegraphics[width=.498\textwidth]{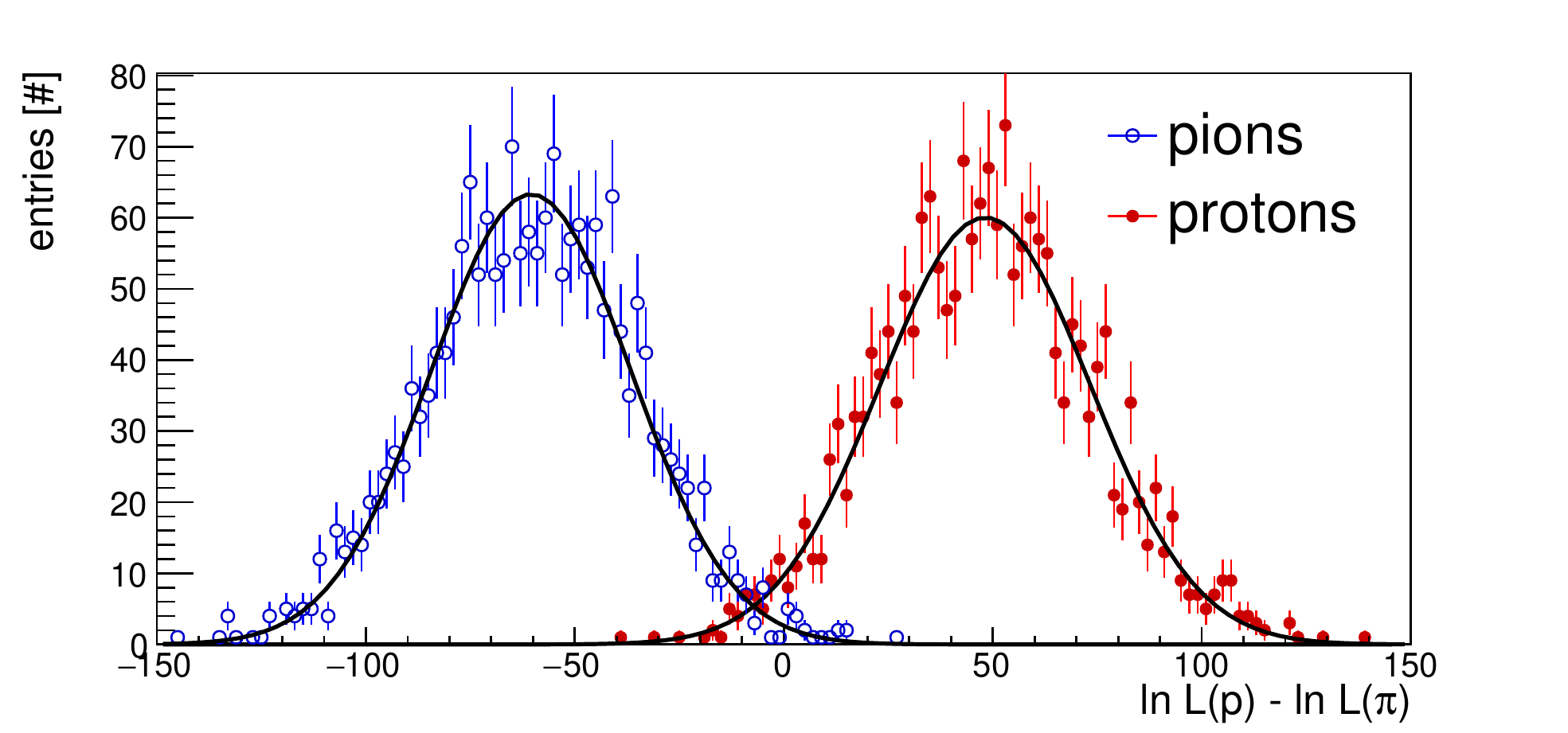}
  \includegraphics[width=.498\textwidth]{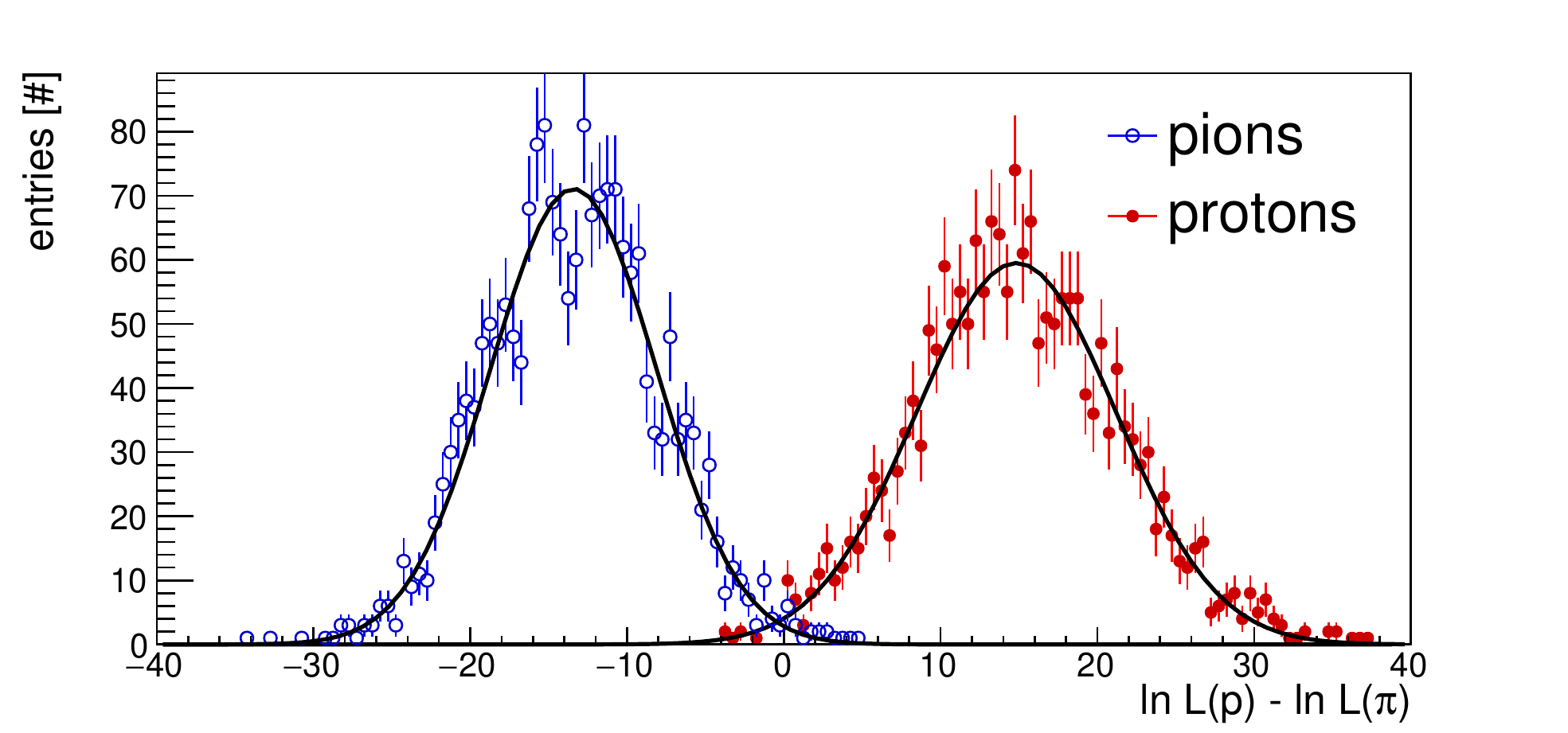}
  \caption{\label{fig:results} The performance of the time imaging reconstruction
    for the PANDA Barrel DIRC prototype simulation.
    The $\pi/p$ log-likelihood difference distributions for
    pions (blue) and protons (red). The $\pi/p$ separation power
    from the Gaussian fits is $4.4 \pm 0.1$ s.d and $4.8 \pm 0.1$ s.d for reconstruction with
    analytical (left) and simulated (right) PDFs, respectively.}
\end{figure}

The resulting likelihood difference distributions of 2k protons and pions
are shown in figure~\ref{fig:results}
for the reconstruction with analytical (left) and simulated (right) PDFs.
The time imaging reconstruction with analytical PDFs
delivers $4.4 \pm 0.1$ s.d. separation which is close to the
$4.8 \pm 0.1$ s.d. obtained with simulated PDFs.
In both cases, the time imaging surpasses the geometrical reconstruction
which delivers $4.1 \pm 0.1$ s.d.

\section{Conclusion}

The time imaging reconstruction uses both position and
time of the detected Cherenkov photons.
The photon propagation time distributions are used to construct probability density
functions for likelihood calculations. The fastest and most efficient
way to create those PDFs is to use analytical calculations.
The initial implementation by the Belle II TOP group was extended
with look-up-tables to account for the specific focusing system
of the PANDA Barrel DIRC. The performance comparison showed that the analytical
approach provides a performance close to the best possible one,
obtained with simulated PDFs. 

\acknowledgments

This work was supported by BMBF, HGS-HIRe, HIC for FAIR.

\end{document}